\documentclass[twocolumn,showpacs,preprintnumbers,amsmath,amssymb]{revtex4}

\usepackage{graphicx}
\usepackage{dcolumn}
\usepackage{bm}
\begin{document}
\preprint{}
\title{Elimination of the Landau pole in QCD with the spontaneously generated anomalous three-gluon interaction}
\author{Boris A. Arbuzov }
\affiliation{Skobeltsyn Institute of Nuclear Physics, Lomonosov Moscow State University\\ Leninskie gory 1, 119991 Moscow, Russia}%
\email{arbuzov@theory.sinp.msu.ru}
\author{Ivan V. Zaitsev}
\affiliation{Skobeltsyn Institute of Nuclear Physics, Lomonosov
Moscow State University\\ Leninskie gory 1,119991 Moscow, Russia}
\date{\today}
\begin{abstract}
We apply the Bogoliubov compensation principle to QCD. The non-trivial solution of compensation equations for a spontaneous generation of the anomalous three-gluon interaction leads to the determination of parameters of the theory, including behavior of the gauge coupling $\alpha_s(Q^2)$ without the Landau singularity, the gluon condensate $V_2\,\simeq\,0.01\,GeV^4$, mass of the lightest glueball $M_G\,\simeq\,1500\,MeV$ in satisfactory agreement with the phenomenological knowledge. The results strongly support the applicability of N.N. Bogoliubov compensation approach to gauge theories of the Standard Model.
\end{abstract}

\maketitle
\section{Introduction}
We are now sure, that QCD is the genuine theory of strong interactions.
In the perturbative region at high momenta the theory excellently describes the totality of data. However at low momenta the perturbative theory fails. Firstly it is seen from the well known momentum dependence of the running coupling. Let us show here the three loop expression for $\alpha_s(\mu)$
\begin{eqnarray}
& &\alpha_s(\mu)\,=\,\frac{4 \pi}{\beta_0 \ln(\mu^2/\Lambda^2)}\biggr[1-\frac{2 \beta_1\,\ln(\ln(\mu^2/\Lambda^2))}{\beta_0^2\,\ln(\mu^2/\Lambda^2)}+
\label{eq:alphas3}\\
& &\frac{4 \beta_1^2}{\beta_0^4\,\ln^2(\mu^2/\Lambda^2)}\biggl(\Bigl(\ln(\ln(\mu^2/\Lambda^2))
-\frac{1}{2}\Bigr)^2+\frac{\beta_2 \beta_0}{8 \beta_1^2}-\frac{5}{4}\biggr)\biggr]\,;\nonumber
\end{eqnarray}
where $\Lambda$ is the QCD scale parameter and
\begin{eqnarray}
& &\beta_0\,=\,11-\frac{2\,N_f}{3}\,;\quad \beta_1\,=\,51-\frac{19\,N_f}{3}\,;
\label{eq:bi}\\
& &\beta_2\,=\,2857-\frac{5033\,N_f}{9}+\frac{325\,N_f^2}{27}\,; \nonumber
\end{eqnarray}
For low momenta region we take expression~(\ref{eq:alphas3}) with number of flavors $N_f\,=\,3$ and take for normalization its value at mass of $\tau$-lepton. We have
\begin{equation}
\alpha_{s}(M_\tau = 1777\,MeV)\,=\,0.32\pm 0.05\,. \label{eq:alphatau}
\end{equation}
From here we obtain
\begin{equation}
\Lambda_{3}\,=\,(345 \pm 19)\,MeV\,. \label{eq:Lambda3}
\end{equation}
Thus from~(\ref{eq:alphas3}) we see, that for $\mu = \Lambda_3$ we have the pole  (and the cut at the same point).

At first such pole was disclosed in QED~\cite{LP,LNB} and thus was called the  Landau pole. The existence of the pole makes a theory internally contradictory.
As for QED, L.D. Landau himself in the issue dedicated to Niels Bohr~\cite{LNB} had first stated, that for a realistic number of the charged elementary fields the pole was situated far beyond the Planck mass and so it presumably could be removed by quantum gravitation effects.
However in QCD pole is situated in the observable region of few hundreds MeV.
As far as we know, there is no way to get rid of such pole in the framework of the perturbation theory. It is a general belief, that non-perturbative contributions somehow exclude the pole. For reviews of different possibilities see {\it e.g.}~\cite{Fischer,SS2}.

In the present work we would demonstrate just how the pole in~(\ref{eq:alphas3}) could be eliminated in the approach to non-perturbative effects in gauge theories, which was induced by the famous N.N. Bogoliubov compensation approach~\cite{Bog1,Bog2}.
\section{ The compensation equation}
For the beginning we consider pure gluon QCD without quarks.
We start with Lagrangian
with gauge group $SU(3)$. That is we define the gauge sector to
be color octet of gluons $F^a_\mu$.
\begin{eqnarray}
& & L\,=\,-\frac{1}{4}\, F_{\mu\nu}^a F_{\mu\nu}^a;\;\label{eq:initial}\\
& &F_{\mu\nu}^a\,=\,
\partial_\mu F_\nu^a - \partial_\nu F_\mu^a\,+g\,f_{abc}F_\mu^b F_\nu^c\,.\nonumber
\end{eqnarray}
where we use the standard notations.
Let us consider a possibility of spontaneous generation of  the following effective interaction
\begin{equation}
-\,\frac{G}{3!}\cdot \,f_{abc}\,
F_{\mu\nu}^a\,F_{\nu\rho}^b\,F_{\rho\mu}^c\,;\label{eq:effint}
\end{equation}
which is
usually called the anomalous three-gluon interaction.

Here notation
 $\frac{G}{3!}\cdot \,f_{abc}\,
F_{\mu\nu}^a\,F_{\nu\rho}^b\,F_{\rho\mu}^c$ means corresponding
non-local vertex in the momentum space
\begin{eqnarray}
& &(2\pi)^4 G\,f_{abc} (g_{\mu\nu} (q_\rho pk - p_\rho qk)+ g_{\nu\rho}
(k_\mu pq - q_\mu pk)+\nonumber\\
& &+g_{\rho\mu} (p_\nu qk - k_\nu pq)+q_\mu k_\nu p_\rho - k_\mu p_\nu q_\rho)\,\times\nonumber\\
& &\times\, F(p,q,k)\,
\delta(p+q+k)\,+...\;;\label{eq:vertex}
\end{eqnarray}
where $F(p,q,k)$ is a form-factor and
$p,\mu, a;\;q,\nu, b;\;k,\rho, c$ are respectively incoming momenta,
Lorentz indices and color indices of gluons.

In accordance to the Bogoliubov approach~\cite{Bog1, Bog2} in application to
QFT~\cite{Arb04} we look for
a non-trivial solution of a
compensation equation, which is formulated on the basis
of the Bogoliubov procedure {\bf add -- subtract}. Namely
let us write down the initial expression~(\ref{eq:initial})
in the following form
\begin{eqnarray}
& &L\,=\,L_0\,+\,L_{int}\,;\nonumber\\
& &L_0\,=\,-\,\frac{1}{4}\,F_{\mu\nu}^a F_{\mu\nu}^a\,+
\frac{G}{3!}\cdot\,f_{abc}\,F_{\mu\nu}^a\,F_{\nu\rho}^b\,F_{\rho\mu}^c\,;
\label{eq:L0}\\
& &L_{int}\,=\,-\,\frac{G}{3!}\cdot\,f_{abc}\,
F_{\mu\nu}^a\,F_{\nu\rho}^b\,F_{\rho\mu}^c\,.\label{eq:Lint}
\end{eqnarray}
Here notation
 $-\,\frac{G}{3!}\cdot \,f_{abc}\,
F_{\mu\nu}^a\,F_{\nu\rho}^b\,F_{\rho\mu}^c$ is already explained~(\ref{eq:vertex}).

We mean also that there are present four-gluon, five-gluon and
six-gluon vertices according to expression for $F_{\mu\nu}^a$
(\ref{eq:initial}). Note, that inclusion of total gluon term $F_{\mu \nu}^a\,F_{\mu \nu}^a$ in the new free Lagrangian~(\ref{eq:L0}) is performed in view of maintaining the gauge invariance of the approach.

Effective interaction~(\ref{eq:effint}) is
called {\bf anomalous three-gluon interaction}. Our interaction constant $G$ is to be defined by the subsequent studies.

Let us consider  expression~
(\ref{eq:L0}) as the new {\bf free} Lagrangian $L_0$,
whereas expression~(\ref{eq:Lint}) as the new
{\bf interaction} Lagrangian $L_{int}$. It is important to note, that we
put into the new {\bf free} Lagrangian the full quadratic in $F$ term including
gluon self-interaction, because we prefer to maintain gauge invariance of the approximation being used. Indeed, we shall use both four-gluon term from the last term
in~(\ref{eq:L0}) and triple one from the last but one term of~(\ref{eq:L0}).
Then compensation conditions (see for details~\cite{Arb04}) will
consist in demand of full connected three-boson vertices of the structure~(\ref{eq:vertex}),
following from Lagrangian $L_0$, to be zero. This demand
gives a non-linear equation for form-factor $F$.

Such equations according to terminology of works
~\cite{Bog1, Bog2} are called {\bf compensation equations}.
In a study of these equations it is always evident the
existence of a perturbative trivial solution (in our case
$G = 0$), but, in general, a non-perturbative
non-trivial solution may also exist. Just the quest of
a non-trivial solution inspires the main interest in such
problems. One can not succeed in finding an exact
non-trivial solution in a realistic theory, therefore
it is of great importance to choose an adequate
approach, the first non-perturbative approximation of
which describes the main features of the problem.
Improvement of a precision of results is to be achieved
by corrections to the initial first approximation.

Thus our task is to formulate the first approximation.
Here the experience acquired in the course of performing
works~\cite{Arb04, Arb05, AVZ} could be helpful. Now in view of
obtaining the first approximation we would make the following
assumptions.\\
1) In compensation equation we restrict ourselves by
terms with loop numbers 0, 1.\\
2) We reduce thus obtained non-linear compensation equation to a linear
integral equation. It means that in loop terms only one vertex
contains the form-factor, being defined above, while
other vertices are considered to be point-like. In
diagram form equation for form-factor $F$ is presented
in Fig.1. Here four-leg vertex correspond to interaction of four
bosons due to our effective three-gluon interaction. In our approximation we
take here the vertex with interaction constant proportional
to $g\,G$.\\
3) We integrate by angular variables of the 4-dimensional Euclidean
space. The necessary rules are presented in paper~\cite{Arb05}.

Let us note that such approximation was previously used in works~\cite{Arb05, AVZ, AVZ2} in the study of spontaneous generation of effective Nambu -- Jona-Lasinio interaction.
It was shown in the works that the results agree with data
with average accuracy $\simeq 10 - 15\%$. Thus we could hope for such accuracy in the present problem.
Let us formulate compensation equations in this
approximation.

For {\bf free} Lagrangian $L_0$ full connected
three-boson vertices with Lorentz structure~(\ref{eq:vertex}) are to vanish. One can succeed in
obtaining analytic solutions for the following set
of momentum variables (see Fig.1): left-hand legs
have momenta  $p$ and $-p$, and a right-hand leg
has zero momenta.
However in our approximation we need form-factor $F$ also
for non-zero values of this momentum. We look for a solution
with the following simple dependence on all three variables
\begin{equation}
F(p_1,\,p_2,\,p_3)\,=\,F\Bigl(\frac{p_1^2\,+\,p_2^2\,+\,p_3^2}{2}\Bigr)\,;\label{eq:123}
\end{equation}
Really, expression~(\ref{eq:123}) is symmetric and it turns to $F(x)$
for $p_3=0,\,p_1^2\,=\,p_2^2\,=\,x$. We consider the representation~(\ref{eq:123})
to be the first approximation and we it would be advisable to take into account the corresponding correction in forthcoming studies. We shall also discuss below some possible corrections due to this problem.
\begin{widetext}
At first let us present the expression for four-boson vertex
\begin{eqnarray}
& &\frac{V(p,m,\lambda;\,q,n,\sigma;\,k,r,\tau;\,l,s,\pi)}{\imath\,(2\,\pi)^4} =  g G \Bigl(f^{amn}
f^{ars}\bigl(U(k,l;\sigma,\tau,\pi,\lambda)-U(k,l;\lambda,\tau,\pi,\sigma)-
U(l,k;\sigma,\pi,\tau,\lambda)+\nonumber\\
& &
U(l,k;\lambda,\pi,\tau,\sigma)+U(p,q;\pi,\lambda,\sigma,\tau)-U(p,q;\tau,
\lambda,\sigma,\pi)-U(q,p;\pi,\sigma,\lambda,\tau)
+U(q,p;\tau,\sigma,\lambda,\pi)\bigr)+\nonumber\\
& &f^{arn}\,
f^{ams}\bigl(U(p,l;\sigma,\lambda,\pi,\tau)-U(l,p;\sigma,\pi,\lambda,\tau)
-U(p,l;\tau,\lambda,\pi,\sigma)+U(l,p;\tau,\pi,\lambda,\sigma)+
U(k,q;\pi,\tau,\sigma,\lambda)-\label{eq:four}\\
& &U(q,k;\pi,\sigma,\tau,\lambda)-U(k,q;\lambda,\tau,\sigma,\pi)
+U(q,k;\lambda,\sigma,\tau,\pi)\bigr)-f^{asn}\,
f^{amr}\bigl(U(k,p;\sigma,\tau,\lambda,\pi)-U(p,k;\sigma,\lambda,\tau,\pi)
+\nonumber\\
& &U(p,k;\pi,\lambda,\tau,\sigma)-U(k,p;\pi,\tau,\lambda,\sigma)-
U(l,q;\tau,\pi,\sigma,\lambda)+
U(l,q;\lambda,\pi,\sigma,\tau)
-U(q,l;\lambda,\sigma,\pi,\tau)+U(q,l;\tau,\sigma,\pi,\lambda)\bigr)\Bigr)
\,;\nonumber\\
& &U(k,l;\sigma,\tau,\pi,\tau)=\bigl(k_\sigma\,l_\tau\,g_{\pi\lambda}-
k_\sigma\,l_\lambda\,g_{\pi\tau}+k_\pi\,l_\lambda\,g_{\sigma\tau}-
(kl)g_{\sigma\tau}g_{\pi\lambda}\bigr)\times F(k,\,l,\,-(k+l))\,.\nonumber
\end{eqnarray}
Here triad $p,\,m,\,\lambda$ {\it etc} means correspondingly incoming momentum, color
index, Lorentz index of a gluon and $F$ is the same form-factor as in expression~(\ref{eq:vertex}).

\begin{figure}
\includegraphics[width=18cm]{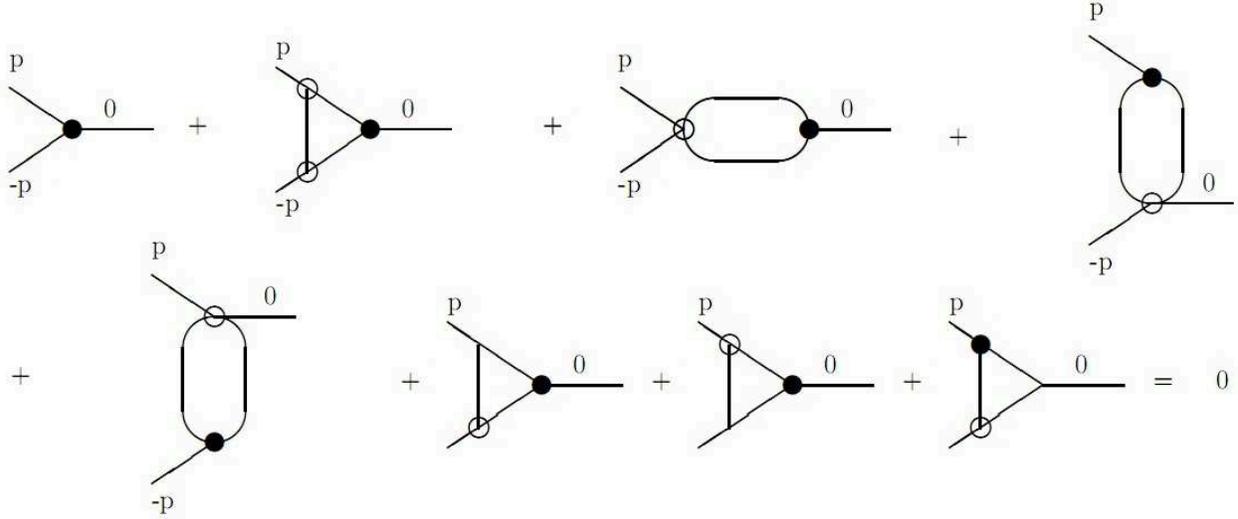}
\caption{Diagrams, describing the compensation equation.
Lines correspond to gluons, black circles correspond to vertex~(\ref{eq:vertex}), open circles to the same vertex with unity form-factor, open circles with four legs correspond to vertex~(\ref{eq:four}), simple point corresponds to usual perturbative vertex.}
\label{fig:Fig1}
\end{figure}

Now according to the rules being stated above we
obtain the following equation for form-factor $F(x)$, which corresponds to Fig.1.
\begin{eqnarray}
& &F(x)\,=\,-\,\frac{G^2\,N}{64\,\pi^2}\Biggl(\int_0^Y\,F(y)\,y dy\,-
\frac{1}{12\,x^2}\,\int_0^x\,F(y)\,y^3 dy\,
+\,\frac{1}{6\,x}\,\int_0^x\,F(y)\,
y^2 dy\,+\frac{x}{6}\,\int_x^Y\,F(y)\,dy\,-\label{eq:F}\\
& &\frac{x^2}{12}\,
\int_x^Y\,\frac{F(y)}{y}\,dy \Biggr)\,+\frac{G\,g\,N}{16\,\pi^2}\,
\int_0^Y F(y) dy + \frac{G g N}{24\,\pi^2} \Biggl(\int_{x}^Y \frac{(5 x- 6 y)}{(x-2 y)}F(y) dy +\int_{\frac{3 x}{4}}^{x} \frac{(3 x-  4 y)^2 (2 y -3 x)}{x^2 (x-2 y)}F(y)
dy\Biggr)\,+\nonumber\\
& &\frac{G g N}{32 \pi^2}\Biggl(\int_x^Y \frac{3(x^2-2 y^2)}{8(2 y-x)^2} F(y) dy + \int_{\frac{3 x}{4}}^{x} \frac{3(4 y-3 x)^2(x^2-4 x y+2 y^2)}{8 x^2(2 y-x)^2} F(y) dy + \int_0^x\frac{5 y^2-12 x y}{16 x^2} F(y) dy +\nonumber\\ & &\int_x^Y\frac{3 x^2- 4 x y - 6 y^2}{16 y^2} F(y) dy\Biggr)\,.\nonumber
\end{eqnarray}
\end{widetext}

Here $x = p^2$ and $y = q^2$, where $q$ is an integration momentum, number of colors $N=3$. Here we also divide the initial equation by coupling constant $G$ in view of looking for non-trivial solutions. Of course, the trivial solution $G\,=\,0$ is always possible.

The last four terms in brackets represent diagrams with one usual gauge vertex (see three last
diagrams at Fig.1) These terms maintain the gauge invariance of results in this approximation. Note that one can additionally check the gauge invariance by introduction of  longitudinal term $d_l\,k_\mu k_\nu/(k^2)^2$ in boson propagators to verify independence of results on $d_l$ in this approximation. Ghost contributions also give zero result in the present approximation due to vertex~(\ref{eq:vertex}) being transverse:
\begin{eqnarray}
& &p_\mu V(p,q,k)_{\mu\nu\rho} = q_\nu V(p,q,k)_{\mu\nu\rho}= k_\rho V(p,q,k)_{\mu\nu\rho}=0 ;\nonumber\\
& &V(p,q,k)_{\mu\nu\rho}=g_{\mu\nu} (q_\rho pk - p_\rho qk)+ g_{\nu\rho}
(k_\mu pq -\label{eq:trans}\\
& & q_\mu pk)+g_{\rho\mu} (p_\nu qk - k_\nu pq)+q_\mu k_\nu p_\rho - k_\mu p_\nu q_\rho\,.\nonumber
\end{eqnarray}
Gauge invariance might be also violated by terms arising from momentum dependence of form-factor $F$. However this problem does not arise in the approximation corresponding to equation~(\ref{eq:F}) and becomes essential for taking into account of $g^2$ terms. In this case ghost contributions also do not cancel. The problem of gauge invariance of the next approximations has to be considered in future studies.

We introduce in equation~(\ref{eq:F})
an effective cut-off $Y$, which bounds a "low-momentum" region where
our non-perturbative effects act
and consider the equation at interval $[0,\, Y]$ under condition
\begin{equation}
F(Y)\,=\,0\,; \label{eq:Y0}
\end{equation}
and for $x\,>\,Y$ we continuously transit to the trivial solution
$G\,=\,0$.
We shall solve equation~(\ref{eq:F}) by iterations. That is we
expand its terms being proportional to $g$ in powers of $x$ and
take at first only constant term. Thus we have
\begin{eqnarray}
& &F_0(x) = - \frac{G^2 N}{64\,\pi^2}\Biggl(\int_0^Y F_0(y) y dy + \frac{x}{6}\,\int_x^Y\,F_0(y)\,dy-\nonumber\\
& &\,\frac{x^2}{12}\,
\int_x^Y\,\frac{F_0(y)}{y}\,dy
 +\frac{1}{6\,x}\,\int_0^x\,F_0(y)\,
y^2 dy\,\,-\label{eq:F0}\\
& &\frac{1}{12 x^2} \int_0^x F_0(y) y^3 dy \Biggr) + \frac{87 G g N}{512\,\pi^2}\,\int_0^Y\,F_0(y)\, dy\,.\nonumber
\end{eqnarray}
Expression~(\ref{eq:F0}) provides an equation of the type which were
studied in papers~\cite{Arb04, Arb05, AVZ},
where the way of obtaining
solutions of equations analogous to (\ref{eq:F0}) are described.
Indeed, by successive differentiation of Eq.(\ref{eq:F0}) we come to
Meijer differential equation~\cite{be}
\begin{eqnarray}
& &\biggl(x\,\frac{d}{dx} + 2\biggr)\biggl(x\,\frac{d}{dx} + 1\biggr)\biggl(x\,\frac{d}{dx} - 1\biggr)\biggl(x\,\frac{d}{dx} - 2\biggr)\times\nonumber\\
& &F_0(x)\,+
\frac{G^2\,N\,x^2}{64\,\pi^2}\,F_0(x) = \label{eq:difur}\\
& &4 \Biggl(- \frac{G^2\,N}{64\,\pi^2} \int_0^Y F_0(y)
y dy + \frac{87 G g N}{512\,\pi^2}\int_0^Y F_0(y) dy
\Biggr)\,;\nonumber
\end{eqnarray}
which solution looks like
\begin{eqnarray}
& &F_0(x) \equiv \Psi_0(z) =  C_1 G_{04}^{10}\Bigl( z\,|1/2, 1, -1/2, -1\Bigr) +\nonumber\\
& & C_2 G_{04}^{10}\Bigl( z\,|1, 1/2, -1/2, -1\Bigr) - \frac{G N}{128 \pi^2}\times \label{eq:solution}\\
& & G_{15}^{31}\Bigl( z\,|^0_{1, 1/2, 0, -1/2, -1}\Bigr) \int_0^Y \Biggl(G \,y\,-\,\frac{87\, g}{8}\Biggr)F_0(y)\,dy\,
;\nonumber\\
& &
z = \frac{G^2\,N\,x^2}{1024\,\pi^2}\,;\nonumber
\end{eqnarray}
where
$$
G_{qp}^{nm}\Bigl( z\,|^{a_1,..., a_q}_{b_1,..., b_p}\Bigr)\,;
$$
is a Meijer function~\cite{be}. In case $q=0$ we write only indices $b_i$ in one
line. Constants $C_1,\,C_2$ are defined by the following boundary conditions
\begin{eqnarray}
& &\Bigl[2\,z^2 \frac{d^3\,\Psi_0(z)}{dz^3}\,+9\,z\,\frac{d^2\,\Psi_0(z)}{dz^2}\,+\,
\frac{d\,\Psi_0(z)}{dz}\Bigr]_{z\,=\,z_0} = 0\,;\nonumber\\
& &\Bigl[2\,z^2\,\frac{d^2\, \Psi_0(z)}{dz^2}\,+5\,z\,\frac{d\, \Psi_0(z)}{dz}\,+\,
\Psi_0(z) \Bigr]_{z\,=\,z_0} = 0\,;\nonumber\\
& & z_0\,=\,\frac{G^2\,N\,Y^2}{1024\,\pi^2}\,.\label{eq:bc0}
\end{eqnarray}

Conditions~(\ref{eq:Y0}, \ref{eq:bc0}) defines set of
parameters
\begin{equation}
z_0\,=\,\infty\,; \quad C_1\,=\,0\,
; \quad C_2\,=\,0\,.\label{eq:z0C}
\end{equation}
The normalization condition for form-factor $F(0)=1$ here is the following
\begin{equation}
-\,\frac{G^2\,N}{64\,\pi^2}\,\int_0^\infty F_0(y)
\,y
dy + \frac{87\,G\,g\,N}{512\,\pi^2} \int_0^\infty F_0(y)\,dy\, =\,1\,.
\label{eq:norm}
\end{equation}
However the first integral in (\ref{eq:norm}) diverges due to asymptotic
$$
G_{15}^{31}\Bigl( z\,|^0_{1,\,1/2,\,0,\,-1/2,\,-1}\Bigr)\,\to\,
\frac{1}{2\,z}\,, \quad z\,\to\,\infty\,;
$$
and we have no consistent solution. In view of this we consider the next
approximation. We substitute solution (\ref{eq:solution}) with account of~(\ref{eq:norm}) into terms of Eq.~(\ref{eq:F}) being proportional to gauge constant $g$ but the
constant ones and
calculate terms  proportional to $\sqrt{z}$. Now we have bearing in mind the normalization condition
\begin{eqnarray}
& &F(x)\equiv\Psi(z)\,=\,1 + \frac{85\, g\,\sqrt{N} \,\sqrt{z}}{96\,\pi}\Biggl(\ln\,z + 4\,
\gamma + \nonumber\\
& &4\,\ln\,2-\frac{1975}{168} +\frac{1}{2} G_{15}^{31}\Bigl( z_0\,|^0_{0, 0, 1/2, -1, -1/2}\Bigr)\Biggr)   -\nonumber\\
& &\frac{2}{3\,z} \int_0^z \Psi(t)\,t\, dt- \frac{2\,z}{3}\,\int_z^{z_0}\,\Psi(t) \frac{dt}{t} +\label{eq:Fg}\\
& & \frac{4}{3\,\sqrt{z}} \int_0^z \Psi(t)
\sqrt{t}\, dt + \frac{4\,\sqrt{z}}{3} \int_z^{z_0} \Psi(t) \frac{dt}{\sqrt{t}}\,\,;\nonumber
\end{eqnarray}
where $\gamma$ is the Euler constant. We look for solution of (\ref{eq:Fg}) in the form
\begin{eqnarray}
& &\Psi(z) = \frac{1}{2} G_{15}^{31}\Bigl( z\,|^0_{1, 1/2, 0, -1/2, -1}
\Bigr) - \nonumber\\
& &\frac{85\,g \sqrt{N}}{128\,\pi}\,G_{15}^{31}\Bigl( z\,|^{1/2}_{1, 1/2, 1/2, -1/2, -1}\Bigr) +\label{eq:solutiong}\\
& &C_1\,G_{04}^{10}\Bigl( z\,|\frac{1}{2},\,1,\,-\frac{1}{2},\,-1\Bigr)\,+
\,C_2\,G_{04}^{10}\Bigl( z\,|1,\,\frac{1}{2},\,-\frac{1}{2},\,-1\Bigr)\,.\nonumber
\end{eqnarray}
We have also conditions
\begin{eqnarray}
& &1\,+\,8\int_0^{z_0}\,\Psi(z)\,dz\,=\,
\frac{87\,g\,\sqrt{N}}{32\,\pi}\,\int_0^{z_0}\Psi_0(z)\,\frac{dz}{\sqrt{z}}\,;
\nonumber\\
& &\Psi(z_0)\,=\,0\,;\label{eq:pht1}
\end{eqnarray}
and boundary conditions analogous to~(\ref{eq:bc0}). The last
condition~(\ref{eq:pht1}) means smooth transition from the non-trivial
solution to trivial one $G\,=\,0$. Knowing form~(\ref{eq:solutiong}) of
a solution we calculate both sides of relation~(\ref{eq:Fg}) in two
different points in interval $0\,<\,z\,<\,z_0$ and having four
equations for four parameters solve the set. With $N\,=\,3$ we obtain
the following solution, which we use to describe QCD case
\begin{eqnarray}
& &g(z_0)=3.8166\,;\; z_0=0.009553;\;\nonumber\\
& &C_1\,=\,-\,5.19055\,; \; C_2\,=\,5.46167\,.\label{eq:gY}
\end{eqnarray}
We would draw attention to the fixed value of parameter $z_0$. The solution
exists only for this value~(\ref{eq:gY}) and it plays the role of eigenvalue.
As a matter of fact from the beginning the existence of such eigenvalue is
by no means evident. This parameter $z_0$ defines scale appropriate to the solution. That is why we take value of running coupling $g$ in solution~(\ref{eq:gY}) just at this point. Note, that in what follows we always use the notation $F(x)$ for the main form-factor of the approach.

It is worth to note, that there is also another solution of the set of equations, which corresponds to larger value of $z_0 \simeq 9.6$ and smaller value of $g(z_0) \simeq 0.6$ for $N=2$. We apply this solution for an adequate description of non-perturbative contributions to the electro-weak interaction~\cite{AZ11,Arb12,AZ12,AZPR}.

Let us recall that from three-loop expression for $\alpha_s(\mu^2)$~(\ref{eq:alphas3}) with number of flavors $N_f = 3 $   we have normalization of its value at mass of $\tau$-lepton~(\ref{eq:alphatau}).

We normalize the running coupling by condition
\begin{equation}
\alpha_{s}(x_0)\,=\,\frac{g(z_0)^2}{4\,\pi}\,=\,1.15515;\label{eq:alphan}
\end{equation}
where coupling constant $g$ entering in expression
~(\ref{eq:gY}) is just corresponding to this normalization point.  Now from definition of $z$~(\ref{eq:solution}) and value $z_0$~(\ref{eq:gY}) we have
\begin{equation}
G\,=\,\frac{1}{\Lambda_G^2}\,; \quad \Lambda_G\,=\,(264 \pm 7)\,MeV\,.\label{eq:GG}
\end{equation}
Thus we have obtained the definite value for the coupling of the interaction~(\ref{eq:effint}) under discussion.
Typical energy scale around $250\,MeV$ is natural for strong interaction.
It is also worth mentioning the value of the momentum which corresponds to boundary of non-perturbative region $z_0$. From Eqs.(\ref{eq:gY}, \ref{eq:GG}) we have for this momentum
\begin{equation}
p_0\,=\,(630 \pm 18)\,MeV\,.\label{eq:p0}
\end{equation}
Non-perturbative boundary~(\ref{eq:p0}) seems also natural from phenomenological point of view.

We have to bear in mind, of course, that all these results are obtained under chosen approximation. For example, change of form of dependence on three variables in expression~(\ref{eq:123}) leads to some change in constant term in inhomogeneous part of equation~(\ref{eq:Fg}). The coefficient afore the logarithm in its second term does not depend on the form, but the constant one can be changed. It is important to understand how small changes in this term influence results. In view of this we consider additional term $\epsilon$ in the inhomogeneous part of~(\ref{eq:Fg}). Thus we have the following modified expression
\begin{eqnarray}
& &1 + \frac{85 g \sqrt{N} \sqrt{z}}{96\,\pi}\Biggl(\ln z + 4
\gamma + 4 \ln 2- \frac{1975}{168} +\nonumber\\
& &\frac{G_{15}^{31}\Bigl( z_0\,|^0_{0, 0, 1/2, -1, -1/2}\Bigr)}{2} +\epsilon\Biggr) ;\label{eq:nhompart1}
\end{eqnarray}

Let us take example  $\epsilon\,=\,0.13$. In this case instead of~(\ref{eq:gY}) we have
\begin{eqnarray}
& &g(z_0)=3.11587\,;\; z_0=0.0153348;\;\nonumber\\
& &C_1\,=\,-\,4.47289\,; \; C_2\,=\,3.62922\,;\label{eq:gY13}
\end{eqnarray}
that in the same way as for case $\epsilon\,=\,0$ leads to the following parameters
\begin{eqnarray}
& &\alpha_{s}(x_0)\,=\,\frac{g(z_0)^2}{4\,\pi}\,=\,0.7726;\nonumber\\
& &G\,=\,\frac{1}{\Lambda_G^2}\,; \quad \Lambda_G\,=\,(273.5 \pm 7.0)\,MeV\,.\label{eq:GG13}
\end{eqnarray}
Another example  $\epsilon\,=\,0.15$. In this case we have
\begin{eqnarray}
& &g(z_0)=3.03685\,;\; z_0=0.0163105;\;\alpha_{s}(x_0)\,=\,0.7339;\nonumber\\
& &
C_1\,=\,-\,4.37005\,; \; C_2\,=\,3.43372\,;\nonumber\\
& &\quad
G\,=\,\frac{1}{\Lambda_G^2}\,; \quad \Lambda_G\,=\,(276.4 \pm 7.0)\,MeV\,.\label{eq:gY15}
\end{eqnarray}
\section{Running coupling}
In previous sections
N.N. Bogoliubov compensation principle~\cite{Bog1, Bog2}
was applied to studies of a spontaneous generation of effective non-local interaction~(\ref{eq:effint}) in QCD.

It is of the utmost interest to study an influence of interaction~(\ref{eq:effint}) on the behavior of strong running coupling $\alpha_s(k^2)$ in the region below $z_0$ {\it i.e.} $k < p_0$(\ref{eq:p0}).

\begin{figure}
\begin{picture}(20,0)
\put(-35,-15){$k$}
\put(55,-30){$k$}
\end{picture}
\includegraphics[width=8cm]{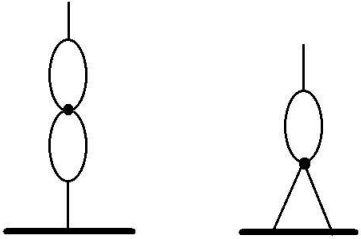}

\caption{Diagrams, describing the contribution of non-perturbative vertex~(\ref{eq:four}), denoted by the black spot, to the running coupling $\alpha_s(k^2)$.
Simple lines correspond to gluons and thick lines correspond to quarks.}
\label{fig:Fig2}
\end{figure}

For the purpose we rely on considerations connected with the renormalization group approach~\cite{BogSh} (for application to QCD see, e.g~\cite{PS}).
We have the one loop perturbative expression for QCD $\beta$-function.
\begin{equation}
\beta(g)\,=\,-\,\frac{g^3}{(4\,\pi)^2}\biggl(11\,-\,\frac{2\,N_f}{3}\biggr)\,;
\label{eq:betapert}
\end{equation}
We shall take additional contributions for small momentum $k^2 \to 0$ of our new interactions according to diagrams shown in Fig. 2 that gives instead of~(\ref{eq:betapert})
\begin{equation}
\beta(g) = - \frac{g^3}{(4 \pi)^2}\biggl[\biggl(11 - \frac{2 N_f}{3}
\biggr) - \frac{405\sqrt{3}\,g(z0)}{2\,\pi}\,\Phi(0)\biggr];\label{eq:beta0}
\end{equation}
where $\Phi(0)$ is the result of calculation of diagrams Fig. 2 (see below).
Here we see a decisive difference in behavior of perturbative $\beta$ ~(\ref{eq:betapert}), which acts at large momenta $k > p_0$ and non-perturbative one for small $k\simeq 0$ ~(\ref{eq:beta0}). According to calculation of $\Phi(0)$ with account of~(\ref{eq:gY}) the sign of $\beta$ changes between these regions. So $\alpha_s(k^2)$ for $k^2 \to 0$ is also positive as well as for large $k$.
To consider a behavior in between we return to definition of the $\beta$-function~\cite{PS}
\begin{eqnarray}
& &\beta(g,t) = g\, M\frac{\partial}{\partial M}\bigl(\delta_{pert}+\delta_{nonpert}\bigr)=\frac{g^3}{(4\pi)^2}\times\label{eq:betaf}\\
& &M \frac{\partial}{\partial M}\frac{\Gamma(2-\frac{d}{2})}{2(M^2)^{2-\frac{d}{2}}}\Bigl(\bigl(11-\frac{2 N_f}{3}\bigr) -\frac{405\sqrt{3}\,g(z_0)}{2\,\pi}\,\Phi(t)\Bigr);\nonumber
\end{eqnarray}
where function $\Phi(t)$ is defined by calculation of diagrams Fig.2 and $d \to 4$ is the space-time dimension.

\begin{eqnarray}
& &\Phi(t) = \int_t^{z_{01}}\frac{u-3 t/4}{u-t/2}F(u) du + \nonumber\\
& &\int_{3 t/4}^t\frac{4(u-3 t/4)^2}{t(u-t/2)}F(u) du ;\; t < z01 ;\label{eq:betanpert}\\
& &\Phi(t) = \int_{3 t/4}^{z_{01}}\frac{4(u-3 t/4)^2}{t(u-t/2)}F(u) du;\, z_{01} < t < \frac{4 z_{01}}{3}\,;\nonumber\\
& &\Phi(t)\,=\,0\,;\; t\,>\,\frac{4 z_{01}}{3}\,;\; z_{01}\,=\,\sqrt{z_0} ;\nonumber\\
& & u=\frac{\sqrt{3}G\,q^2}{32\,\pi}\,;\; t=\frac{\sqrt{3}G\,k^2}{32\,\pi}\,.\nonumber
\end{eqnarray}
This leads to modification of relation~(\ref{eq:beta0})
\begin{equation}
\beta(g,t) = - \frac{g^3}{(4 \pi)^2}\biggl[\bigl(11 - \frac{2 N_f}{3}
\bigr) - \frac{405\sqrt{3}\,g(z0)}{2\,\pi} \Phi(t)\biggr];\label{eq:beta}
\end{equation}

Thus in approximation using the
two-loop expression corresponding to diagrams of Fig.2  we have for $N_f=3$
\begin{equation}
\alpha_s(k^2) = \frac{\alpha(k^2_0)}{1 + \frac{\alpha_s( k^2_0)}{6 \pi}\Bigl(\frac{27}{2}-\frac{405\sqrt{3} g(z0)}{2\pi}\Phi(t)\Bigr)\ln \frac{k^2}{k^2_0}} \,;\label{eq:alphax}
\end{equation}
where $t$ is defined in~(\ref{eq:betanpert}).
With $G$ defined by~(\ref{eq:GG}), $g(z_0)$ defined by~(\ref{eq:gY}) and $k^2=Q^2$ we have the behavior of $\alpha_s(Q)$.
With fixed parameter $\epsilon$ in~(\ref{eq:nhompart1}) we calculate the behavior of running coupling. Let us begin with initial case $\epsilon\,=\,0$.
We have value of $\alpha_s(Q)$ at the beginning point of non-perturbative contribution, corresponding to $\bar z_{01}=\frac{4}{3}z_{01}$, corresponding to momentum $Q = 726\,MeV$.
\begin{equation}
\alpha_s(\bar z_{01})\,=\,0.936\,. \label{eq:nalpha}
\end{equation}
The boundary of non-perturbative region $Q_0 = 726\,MeV$ seem quite reasonable.

Now the behavior of $\alpha_s(Q)$ is drawn in Fig.3. We would like to draw attention to the result, presented at Fig.3 , which consists in absence of Landau pole in expression~(\ref{eq:alphax}). Remind, that in perturbative calculation up to four loops the singularity at Landau pole point is always present. Only by taking into account of the non-perturbative effects we achieve elimination of this very unpleasant feature, which was seriously considered as a sign of the inconsistency of the quantum field theory~\cite{LP, LNB}.

\begin{figure}
\includegraphics[width=6.8cm]{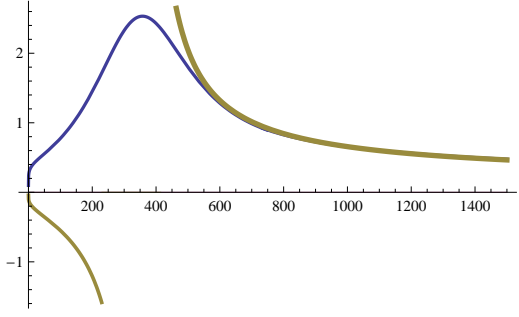}

\caption{Dependence of the running coupling $\alpha_s(Q),\,Q$ in MeV, with $\epsilon=0$.
The continuous line corresponds to $\alpha_s$ with non-perturbative contribution~(\ref{eq:alphax}), the discontinuous one with a pole corresponds to the usual perturbative one-loop expression.}
\label{fig:Fig3}
\end{figure}
\begin{figure}
\begin{picture}(20,10)
\end{picture}
\includegraphics[width=6.8cm]{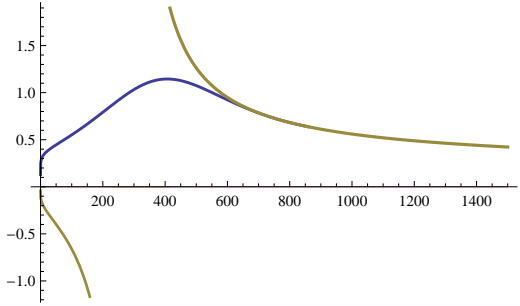}

\caption{Dependence of the running coupling $\alpha_s(Q)$ for $\epsilon=0.13$.
The continuous line corresponds to $\alpha_s$ with non-perturbative contribution~(\ref{eq:alphax}), the discontinuous one with a pole corresponds to the usual perturbative one-loop expression.}
\label{fig:Fig4}
\end{figure}

There is also a feature of expression~(\ref{eq:alphax}), which deserves being mentioned. The limit of $\alpha_s(Q)$ for $Q \to 0$ is zero. Such possibility is also discussed on phenomenological grounds. In particular, there are indications for decreasing of $\alpha_s(Q^2)$ for $Q^2\,\to\,0$ in studies of low-mass resonances~\cite{ItSh}. Let us note, that a number of lattice calculations of
the running coupling also give a similar behavior~\cite{SKW,PhB,Berlin,KF,BB}.

Let us also consider $\alpha_s$ behavior for other values of parameter $\epsilon$. The behavior for $\epsilon=0.13$ is presented in Fig.4. The pole here is also absent, but values of $\alpha_s$ in non-perturbative region are smaller than for case $\epsilon=0$.
The average $\alpha_s$ in the non-perturbative region for $\epsilon=0.13$
\begin{eqnarray}
& &\bar \alpha_s\,=\,\frac{1}{Q_0}\int_0^{Q_0}\,\alpha_s(Q)\,dQ\,=\,0.87\,;\nonumber\\ & &Q_0\,=\,\sqrt{\frac{128 \pi z_{01}}{3 \sqrt{3}\,G}}\,.\label{eq:alphaaver13}
\end{eqnarray}
For $\epsilon=0.15$ $\bar \alpha_s\,=\,0.84$.

\section{the gluon condensate}
One of important non-perturbative parameters is the gluon condensate, that is the following vacuum average
\begin{equation}
V_2\,=\,<\frac{g^2}{4\,\pi^2}\,F^a_{\mu \nu}\,F^a_{\mu \nu}>\,.\label{eq:V2}
\end{equation}
Let us estimate this parameter in our approach. We apply our method to the first non-perturbative contributions, presented at Fig.5, which is proportional to $g\,G$. It is important to introduce Feynman rule for
contribution of operator~(\ref{eq:V2}) in brackets. We denote it by skew cross in Fig.5
\begin{equation}
V_{FF}(\mu,\nu;p)\,=\,\imath\,\frac{g^2}{\pi^2}\,(g_{\mu \nu}\,p^2 - p_\mu\,p_\nu)\,.\label{eq:FeynmanV2}
\end{equation}

With distribution of integration momenta denoted in Fig. 5 form-factor in both types of diagrams according to~(\ref{eq:123}) has the same argument:
\begin{equation}
F(p^2+\frac{3}{4}q^2)\,.\label{eq:Fpq}
\end{equation}

\begin{figure}
\includegraphics[width=9cm]{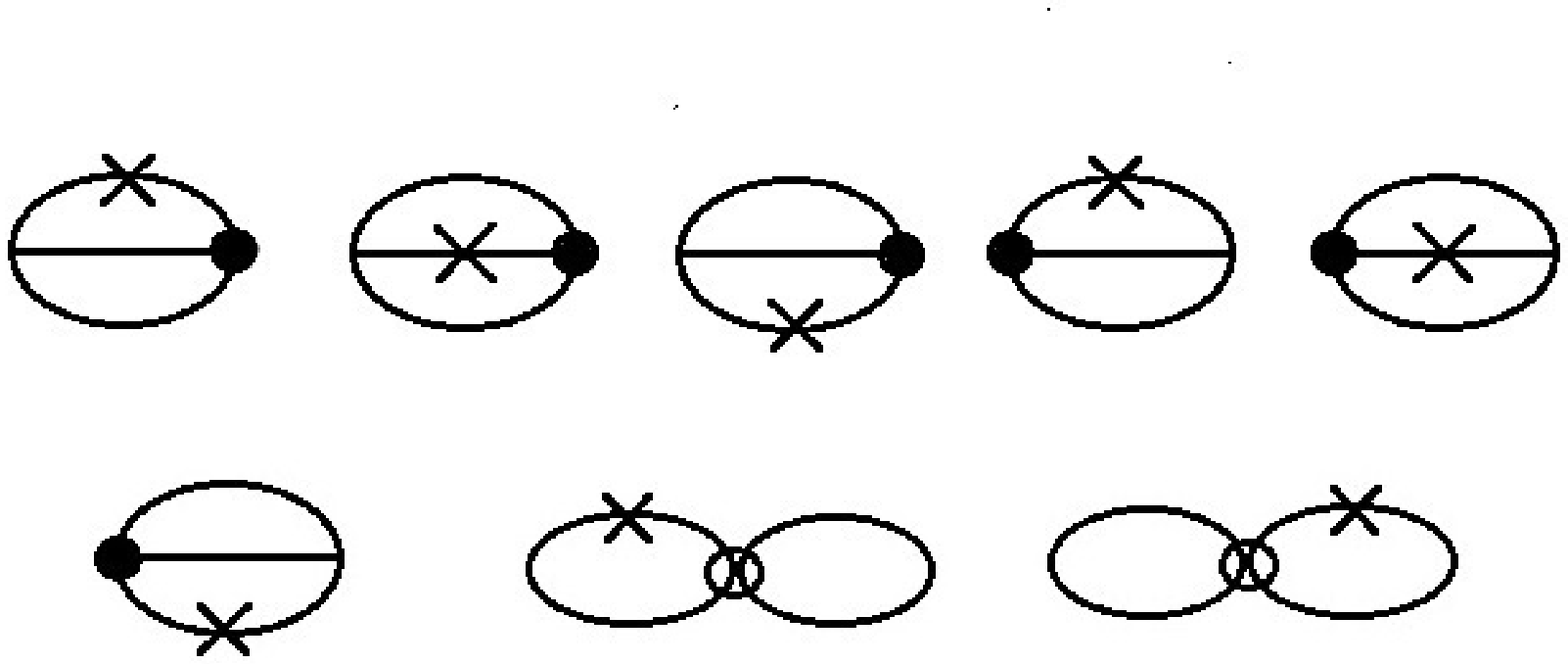}
\caption{Diagrams for calculation of the gluon condensate. Lines -- gluons, black circle -- triple vertex~(\ref{eq:vertex}), open circle -- four gluon vertex~(\ref{eq:four}) with corresponding form-factor and skew cross -- vertex~(\ref{eq:FeynmanV2}). Momenta directed to the right are p-q/2, q, -p-q/2 for bug-like diagrams and p-q/2, p+q/2 for $\infty$-like diagrams.}
\label{fig:Fig5}
\end{figure}

It comes out, that the second and the third terms in the second row of Fig.~\ref{fig:Fig5} are twice each of the previous terms. Thus the sum is equal to the result for the first diagram multiplied by 10.

We have after the Wick rotation
\begin{eqnarray}
& &V_2\,=\,\frac{10\times 24\,g^3\,G}{(2\,\pi)^8\,\pi^2}\times\label{eq:V2Intmom}\\
& &\int F(p^2+\frac{3}{4}q^2)\frac{12\,(p^2\,q^2-pq^2)}{q^2(p-q/2)^2(p+q/2)^2}dp dq\,.\nonumber
\end{eqnarray}
Using the following integral by angle
\begin{eqnarray}
& &\int_0^\pi\frac{\sin^2(\theta)\,d\theta}{(p^2+\frac{q^2}{4})^2-(p q)^2}\,=\,\label{eq:intcos2}\\
& &
\frac{\pi}{2(x+\frac{y}{4})}\Bigl[\theta\bigl(x-\frac{y}{4}\bigr)\frac{1}{x}+
\theta\bigl(\frac{y}{4}-x\bigr)
\frac{4}{y}\Bigr]\,;\nonumber\\
& & x =p^2\,;\quad y=q^2\,;\nonumber
\end{eqnarray}
we obtain the following expression for quantity~(\ref{eq:V2Intmom})
\begin{eqnarray}
& &V_2\,=\,\frac{5\, g^3\,2^{11}}{G^2 \pi^3\,\sqrt{3}}\int_0^{\sqrt{z_0}}F(t)\,I_t\,  dt\,;\nonumber\\
& & I_t\,=\,12\,\biggl(-\int_0^t\frac{(t-y)^2}{t-y/2}dy - \nonumber\\
& &4\int_t^{4 t/3}\frac{(t-y)^2(t-3 y/4)}{(t-y/2)y}dy+\int_0^{4 t/3}\Bigl(t-\frac{3 y}{4}\Bigr)dy\,\biggr)\,;\nonumber\\
& &t\,=\,\frac{G\,\sqrt{3}}{2^5\,\pi}\biggl(x\,+\,\frac{3\,y}{4}\biggr)\,.
\label{eq:intV2}
\end{eqnarray}
We have already expressions~(\ref{eq:solutiong}, \ref{eq:gY}) for form-factor $F(z),\,z=t^2$. So calculation here is direct and we obtain, using values for $g$~(\ref{eq:gY}) and the central value in definition of $G$~(\ref{eq:GG})
\begin{eqnarray}
& &V_2\,=\,\frac{5\, g^3\,2^{10}}{\pi^3 \sqrt{3}\,G^2} 12\Bigl(2-6 \ln \frac{4}{3}\Bigr) \int_0^{z_0} F(z)\sqrt{z}\,dz \,=\nonumber\\
& &0.00955\,GeV^4\,;\label{eq:V2N}
\end{eqnarray}
Provided we take nonzero value for $\epsilon$ in expression~(\ref{eq:nhompart1}) results for gluon condensate  read
\begin{eqnarray}
& &V_2\,=\,0.0120\,GeV^4\,(\epsilon\,=\,0.13);\nonumber\\
& &V_2\,=\,0.0128\,GeV^4\,(\epsilon\,=\,0.15)\,.\label{eq:V2Ne}
\end{eqnarray}
So in this approximation we have the non-zero non-perturbative parameter $V_2$. Its value agrees within accuracy of determination of this parameter with phenomenological values $V_2 \simeq 0.012\, GeV^4$~\cite{SVZ}, $V_2 \simeq 0.010\, GeV^4$~\cite{Zakh}. Values~(\ref{eq:V2N}, \ref{eq:V2Ne}) show variation in the range of uncertainty of its phenomenological definition. Thus we can state, that our non-perturbative approach allows to calculate safely this important parameter.

Let us also estimate vacuum average $V_3$
\begin{equation}
V_3\,=\,<g^3\,f_{a b c}\,F_{\mu \nu}^a\,F_{\nu \rho}^b\,F_{\rho \mu}^c>\,.\label{eq:V3}
\end{equation}
 Quite analogous calculations give {\it e.g.} with $\epsilon\,=\,0.13$
\begin{eqnarray}
& &V_3\,=\,\frac{g^3\, 2^{17}}{ G^3}\Bigl(2 - 6\ln\frac{4}{3}\Bigr)\int_0^{z_0} z F(z) dz\,=\nonumber\\
& &0.00744\, GeV^6\,.\label{eq:V3Ne}
\end{eqnarray}
\section{The glueball}

The existence of anomalous interaction~(\ref{eq:effint}) makes possible to consider gluonic states. We shall consider scalar glueball $X_0$ state to get indications if value of the non-perturbative constant~(\ref{eq:GG}) may be used for adequate description of the non-perturbative effects of the strong interaction. For the purpose we use
Bethe-Salpeter equation with the kernel corresponding to one-gluon exchange with our (point-like) anomalous three-gluon interaction~(\ref{eq:effint}). We take for vertex of $X_0$ interaction with two gluons in the following form
\begin{equation}
\frac{G_{gb}}{2}\,F_{\mu \nu}^a\,F_{\mu \nu}^a\,X_0\,\Psi_{gb}(x)\,;\quad x = p^2\,;\label{eq:FFX0}
\end{equation}
where $\Psi_{gb}(x)$ is a Bethe-Salpeter wave function.
We have for the first approximation (zero momentum of $X_0$)
\begin{eqnarray}
& &\Psi_{gb}(x)= -\frac{3 G^2}{16 \pi^2}\biggl(\frac{1}{2 x^2}\int_0^x y^3 \Psi_{gb}(y)dy-\nonumber\\
& &
\frac{1}{x}\int_0^x y^2 \Psi_{gb}(y)dy-3\int_0^Y y \Psi_{gb}(y)dy -\label{eq:EQgb}\\
& & x \int_x^Y  \Psi_{gb}(y)dy+\frac{x^2}{2}\int_x^Y \frac{\Psi_{gb}(y)}{y}dy\,;\nonumber
\end{eqnarray}
where we take again the upper limit $Y$ of integration as in~(\ref{eq:F}) due to form-factor of interaction~(\ref{eq:effint}) $F(x)=0$ for $x \ge Y$. 
Again by successive differentiations we obtain from Eq.(\ref{eq:EQgb})
the following differential equation
\begin{eqnarray}
& &\biggl(z'\frac{d}{dz'}+1\biggr)\biggl(z'\frac{d}{dz'}+\frac{1}{2}\biggr)
\biggl(z'\frac{d}{dz'}-\frac{1}{2}\biggr)\times\nonumber\\
& &\biggl(z'\frac{d}{dz'}-1\biggr)\Psi_{gb}(z')=
z'\Psi_{gb}(z')+\frac{C}{4}\,;\label{eq:diffgb}\\
& &C=4\int_0^{\bar z_0}\Psi_{gb}(t') dt';\; z'=\frac{9\, G^2 x^2}{128 \pi^2};\;t'=\frac{9\, G^2 y^2}{128 \pi^2}.\nonumber
\end{eqnarray}
Comparing variable $z'$ in Eq.(\ref{eq:diffgb}) with the initial variable $z$ in Eq(\ref{eq:solution}) we see relation $z'=24\, z$. This means also, that
$\bar z_0=24 z_0$,  $z_0$ from solution~(\ref{eq:gY}).
In new variables Eq.(\ref{eq:EQgb}), in which we also have taken into account terms, proportional to gauge coupling $g$ and mass of the bound state squared $m^2$, looks like
\begin{eqnarray}
& &\Psi_{gb}(z')\,=\,1-\frac{2}{3\,z'}\int_0^{z'}\,\Psi_{gb}(t') t' dt'+\nonumber\\
& &\frac{4}{3\,\sqrt{z'}}\int_0^{z'}\,\Psi_{gb}(t')\sqrt{t'} dt'+
\frac{4 \sqrt{z'}}{3}\int_{z'}^{\bar z_0}\,\frac{\Psi_{gb}(t')}{\sqrt{t'}} dt'-\nonumber\\
& &\frac{2 z'}{3}\int_{z'}^{\bar z_0}\,\frac{\Psi_{gb}(t')}{t'} dt'\,;\label{eq:intgb}\\
& &1 = 4\int_0^{\bar z_0} \Psi_{gb}(t') dt' +\biggl(\kappa + \frac{3 g \sqrt{2}}{2 \pi}\biggr) \int_0^{\bar z_0} \frac{\Psi_{gb}(t')}{\sqrt{t'}} dt' .\nonumber
\end{eqnarray}
Here $\kappa$ is connected with the bound state mass $m$ in the following way
\begin{equation}
\kappa\,=\,-\,\frac{3\, G\, m^2}{8 \sqrt{2}\, \pi}\,.\label{eq:kappa}
\end{equation}
According to expression~(\ref{eq:diffgb}) we look for the solution of Eq.(\ref{eq:EQgb}) in the following form
\begin{eqnarray}
& &\Psi_{gb}(z')\,=\,\frac{\pi}{2}\,G_{1 5}^{2 1} \Bigl( z'|^0_{1,\,0,\,1/2,\,-1/2,\,-1}\Bigr)+\label{eq:Psigb}\\
& &C_1\,G_{0 4}^{2 0} \Bigl( z'|1,\,1/2,\,-1/2,\,-1\Bigr)+\nonumber\\
& &C_2\,G_{0 4}^{1 0} \Bigl(- z'|1,\,1/2,\,-1/2,\,-1\Bigr)\,.\nonumber
\end{eqnarray}
By substituting expression~(\ref{eq:Psigb}) into set of equations~(\ref{eq:intgb}) and using the values of $g$ and $z_0$~(\ref{eq:gY}) we obtain unique solution for parameters
\begin{equation}
C_1\,=\,1.07899\,;\; C_2\,=\,-1.38099\,;\; \kappa\,=\,-2.6415\,.\label{eq:gbCkappa}
\end{equation}
Now from values~(\ref{eq:GG}, \ref{eq:gbCkappa}), using relation~(\ref{eq:kappa}), we have the lightest scalar glueball mass
\begin{equation}
m\,=\,1479 \pm 40\,MeV\,.\label{eq:mgb}
\end{equation}
This value is quite natural, the more so, that the most serious candidate for
being the lightest scalar glueball is the state $f_0(1500)$ (see recent review~\cite{Oks}) with mass $1507 \pm 5\, MeV$, that evidently agrees our number~(\ref{eq:mgb}).

Now we have to obtain the coupling constant of the scalar gluon entering in the expression of the effective interaction~(\ref{eq:FFX0}).
For the purpose we use the normalization condition for Bethe-Salpeter wave function $\Psi(t)$.
\begin{equation}
1\,=\,\frac{\sqrt{2}\,G_{gb}^2}{\pi\,G}\,\int_0^{\bar z_0} \frac{\Psi_{gb}(t')^2}{\sqrt{t'}}\,dt'\,.\label{eq:GFFX}
\end{equation}
Substituting into Eq.(\ref{eq:GFFX}) solution~(\ref{eq:Psigb}, \ref{eq:gbCkappa})
and calculating the integral, we obtain

\begin{eqnarray}
& &G_{gb}^2\,=\, \frac{\pi\,G}{\sqrt{2}\,I}\,=\,1.825\, G\,;\nonumber\\
& &I\,=\,\int_0^{\bar z_0} \frac{\Psi_{gb}(t')^2}{\sqrt{t'}}\,dt'\,=\,1.21732.\label{eq:Ggb}
\end{eqnarray}
From result~(\ref{eq:Ggb}) we have the following value of the glueball coupling

\begin{equation}
G_{gb}\,=\,\frac{1}{190.337\,MeV}\,=\,\frac{5.254}{GeV}\,.\label{eq:Ggbnum}
\end{equation}

\section{Conclusion}

An existence of a non-trivial solution of a compensation
equation is extremely
restrictive. In the most cases such solutions do not exist at all. When we start from a
renormalizable theory we have arbitrary value for
its coupling constant. Provided there exists {\it
stable non-trivial solution of a compensation equation} the
coupling is fixed as well as the parameters of this
non-trivial solution. Note, that application of the same approach to the electro-weak theory~\cite{AZ11,Arb12,AZ12,AZPR,Arb09} also leads to strong restrictions on parameters of the theory including the coupling constant.

We also may state, that in the case, discussed in the present paper, just the non-trivial solution is the stable one, because the theory with the Landau pole is unstable.

We consider the results for the gluon condensate~(\ref{eq:V2N},\ref{eq:V2Ne}) and the glueball mass~(\ref{eq:mgb}) as a confirmation of efficiency of our approach in application to non-perturbative contributions to QCD.

Thus we consider the present results for low-momenta $\alpha_s$
to be encouraging and promising for further applications of
the Bogoliubov compensation approach to principal problems
of elementary particles physics.
\\
\section{Acknowledgments}

The work was supported in part by grant of Ministry of education and science of RF No. 8412 and by grant NSh-3920.2012.2.

\end{document}